\documentclass[12pt]{article}
\usepackage{amssymb}
\usepackage[mathscr]{eucal}
\usepackage{epic}
\usepackage{eepic}
\usepackage{amsfonts}
\usepackage{bm}

  \newtheorem{theorem}{Theorem}

\newcommand{\nc}{\newcommand}
\nc{\BBS}{{\rm BBS}\ } \nc{\g}{\gamma}  \nc{\lm}{\lambda}
\nc{\la}{\lambda} \nc{\bh}{{\bf h}}  \nc{\av}{\prod_{s\,\in\,\ZN}}
\nc{\cR}{{\cal R}} \nc{\kp}{{\varkappa}} \nc{\om}{\omega}
\nc{\qt}{\tilde{q}} \nc{\tp}{\tilde{p}} \nc{\rt}{\tilde{r}}
\nc{\ty}{\tilde{y}} \nc{\tx}{\tilde{x}} \nc{\tQ}{{\widetilde Q}}
\nc{\trh}{\tilde{\rho}}\nc{\ny}{\nonumber}\nc{\lk}{\left(}
\nc{\rk}{\right)} \nc{\Rb}{\right]} \nc{\Lb}{\left[}
\nc{\rb}{\right\}} \nc{\lb}{\left\{} \nc{\hs}{\hspace*{1cm}}
\nc{\hx}{\hspace*{3mm}} \nc{\hq}{\hspace*{6mm}} \nc{\JStP}{{\it J. Stat.
Phys.}}  \nc{\IJMP}{{\it Intern. J. Mod. Phys.}}

\nc{\al}{\alpha}  \nc{\ZN}{\mathbb{Z}_N}
\nc{\bg}{\boldsymbol{\gamma}} \nc{\bdr}{\boldsymbol{\rho}}
\nc{\bu}{{\bf u}} \nc{\bv}{{\bf v}} \nc{\bV}{{\bf V}}
\nc{\rnk}{r_{n,k}} \nc{\lns}{\la_{n,s}} \nc{\lnl}{\la_{n,l}}
\nc{\FD}{{\cal F}} \nc{\lnk}{\la_{n,k}} \nc{\xnl}{x_{n,l}}
\nc{\Psr}{\Psi_{\bdr_n}} \nc{\ap}{a_{n+1}} \nc{\bp}{b_{n+1}}
\nc{\cp}{c_{n+1}} \nc{\dpp}{d_{n+1}}
\nc{\bep}{\bu_{n+1}^{-1}(\ap-\bp\bv_{n+1})} \nc{\Xop}{\mathbf{X}}
\nc{\Zop}{\mathbf{Z}}  \nc\s{{\gamma}}
\def\r#1{(\ref{#1})}

\nc{\ra}{\rangle} \nc{\BAR}{\begin{array}} \nc{\EAR}{\end{array}}
\nc{\bdm}{\begin{displaymath}} \nc{\edm}{\end{displaymath}}
\nc{\be}{\begin{equation}} \nc{\ee}{\end{equation}}
\nc{\ba}{\begin{array}} \nc{\ea}{\end{array}}
\nc{\bea}{\begin{eqnarray}} \nc{\eea}{\end{eqnarray}}

\nc\sip{\gamma^\prime}\nc\ma{a'}\nc\mb{b'}\nc\mc{c'}\nc\md{d'}
\nc\pra{a''}\nc\prb{b''}\nc\prc{c''}\nc\prd{d''}
\nc{\hu}{{\bf u}}\nc{\hh}{{\hat h}}\nc{\bl}{\boldsymbol{\lambda}}\nc{\hv}{{\bf v}}
\nc\si{{\mathrm{s}}}

\nc{\tr}{\mbox{tr}\,}

\begin{document}
 \begin{center}
 \LARGE{Eigenvectors of Baxter--Bazhanov--Stroganov\\ $\tau^{(2)}(t_q)$ model with fixed-spin boundary conditions}
 \end{center}
  \begin{center}\large
N.~Z.~Iorgov\footnote{iorgov@bitp.kiev.ua}, V.~N.~Shadura\footnote{shadura@bitp.kiev.ua},
Yu.~V.~Tykhyy\footnote{tykhyy@bitp.kiev.ua}
 \end{center}
  \begin{center}
Bogolyubov Institute for Theoretical Physics, Kiev, Ukraine
 \end{center}

\begin{abstract}
The aim of this contribution is to give the explicit formulas for
the eigenvectors of the transfer-matrix of
Baxter--Bazhanov--Stroganov (BBS) model ($N$-state spin model) with fixed-spin boundary
conditions. These formulas are obtained by a limiting procedure
from the formulas for the eigenvectors of periodic BBS model.
The latter formulas were derived in the framework of the Sklyanin's method of separation of variables.
In the case of fixed-spin boundaries the corresponding $T-Q$ Baxter equations
for the functions of separated variables are solved explicitly. As a particular case we obtain
the eigenvectors of the Hamiltonian of Ising-like $\mathbb{Z}_N$ quantum chain model.
\end{abstract}

\section{Introduction}

During the last two decades, a considerable progress in  application of
the Separation of Variables  (or functional Bethe ansatz) method  to a broad class of  integrable
models of quantum chains and statistical physics has been achieved.
This progress was initiated by the paper
\cite{Skl} of Sklyanin who has proposed  a recipe for a Separation
of Variables  in the case of quantum Toda chain where the algebraic
Bethe ansatz fails. The idea is to write the eigenvectors of periodic problem as a linear
combinations of the eigenvectors of an auxiliary problem (the auxiliary problem for the periodic Toda chain
is the open Toda chain).
The next step is to construct recursively the eigenvectors of the $m$-site auxiliary problem
in terms of the eigenvectors of the $(m-1)$-site auxiliary problem \cite{KhLeb,DKM}.

In \cite{GIPS}, this program  was realized for the inhomogeneous periodic
Baxter--Bazhanov--Stroganov  (BBS) model (or $\tau^{(2)}(t_q)$ model) \cite{B_tau,BaxInv,BS,BBP}
defined in terms of cyclic $L$-operators \cite{Kore,BS}.
It was shown that for every eigenvalue (found from the functional relations) of the transfer-matrix
there corresponds an explicit formula for the eigenvector.
So the problem of finding the eigenvectors reduces to the problem of finding the corresponding eigenvalues.

At special values of parameters,
the periodic BBS model can be interpreted as a model with fixed-spin boundary
conditions \cite{B_tau}. In this paper we derive
the eigenvectors of this model from the eigenvectors of periodic BBS model
by specializing corresponding parameters. In the case of fixed-spin boundary
conditions, the structure of eigenvalues is simple,
so the $T-Q$ Baxter equations
for the functions of separated variables can be solved explicitly
in terms of cyclic function $w_p(\g)$ \cite{BB} which is a root of unity analog of the $q$-gamma function.
It gives us formulas for the eigenvectors of the BBS model with fixed-spin boundary
conditions.

At the end of the paper we show that the Hamiltonian of the Ising-like $\mathbb{Z}_N$ quantum chain
model with fixed boundary spins
found by Baxter \cite{BaxInv,BaxterZN} can be obtained in the present framework by further specialization
of parameters. It means, in particular, that we can find explicit formulas for the eigenvectors and the eigenvalues
of the Hamiltonian. The formula for the eigenvalues coincides with Baxter's one obtained
from the functional relation for the transfer-matrix. Explicit formulas for the eigenvectors prove
that the corresponding eigenvalues enter to the spectrum of the Hamiltonian with the multiplicity one.

\section{Lattice and quantum chain formulations of \BBS model}

\begin{figure}[ht]
\begin{center}
\renewcommand{\dashlinestretch}{30}
\unitlength=0.04pt
\begin{picture}(8622,4539)(0,-10)
\drawline(2715,3612)(5415,3612)(5415,912)
    (2715,912)(2715,3612)
\drawline(15,3612)(2715,3612)(2715,4512)
\drawline(5415,4512)(5415,3612)(8115,3612)
\drawline(15,912)(2715,912)(2715,12)
\drawline(5415,12)(5415,912)(8115,912)
\dashline{60.000}(15,3612)(2715,912)(5415,3612)(8115,912)
\dashline{60.000}(15,912)(2715,3612)(5415,912)(8115,3612)
\put(4600,2217){\makebox(0,0)[cc]{$m_{k-1}$}}
\put(7200,2217){\makebox(0,0)[cc]{$m_k$}}
\put(1900,2217){\makebox(0,0)[cc]{$m_{k-2}$}} \put(3100,462)
{\makebox(0,0)[cc]{$\g_{k-1}$}} \put(5685,462)
{\makebox(0,0)[cc]{$\g_k$}}
\put(3100,4017){\makebox(0,0)[cc]{$\g'_{k-1}$}}
\put(5730,4017){\makebox(0,0)[cc]{$\g'_k$}} \put  (15,912)
{\makebox(0,0)[cc]{$\bullet$}}
\put(2715,3612){\makebox(0,0)[cc]{$\bullet$}} \put(2715,912)
{\makebox(0,0)[cc]{$\bullet$}}
\put(5415,3612){\makebox(0,0)[cc]{$\bullet$}} \put(5415,912)
{\makebox(0,0)[cc]{$\bullet$}}
\put(8115,3612){\makebox(0,0)[cc]{$\bullet$}} \put(8115,912)
{\makebox(0,0)[cc]{$\bullet$}} \put  (15,3612)
{\makebox(0,0)[cc]{$\bullet$}}
\put(1365,2262){\makebox(0,0)[cc]{$\bullet$}}
\put(4065,2262){\makebox(0,0)[cc]{$\bullet$}}
\put(6765,2262){\makebox(0,0)[cc]{$\bullet$}}
\end{picture}
\end{center}
\caption{\footnotesize{The triangle with vertices marked by the spin
variables $\g_{k-1}$, $\g'_{k-1}$, $m_{k-1}$ corresponds to the
function $F'_{k-1}(\g_{k-1}-\g'_{k-1},m_{k-1})$ in (\ref{bbs_weight});
the triangle $\g_{k}$, $\g'_{k}$, $m_{k-1}$ to
$F''_{k}(\g_{k}-\g'_{k},m_{k-1})$. }} \label{tri}\end{figure}
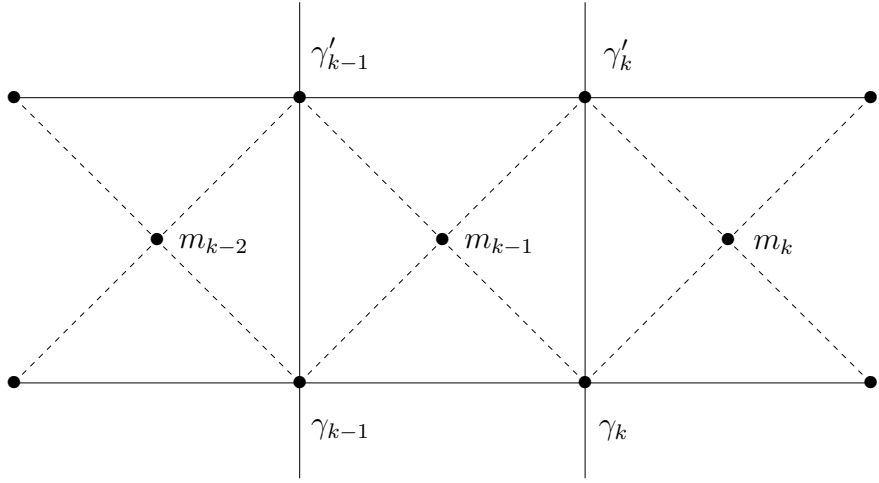

Following the notation of a recent paper of Baxter \cite{B_tau}, we define the BBS model
as a statistical model of short-range interacting spins placed at the vertices of a
rectangular lattice. We label the spin variables
$\si_{x,y}$  by a pair $(x,y)$ of integers: $x=1,\ldots,n+1,$ and
$y=1,\ldots,m $. Each spin variable $\si_{x,y}$ takes  $N$ values ($N\ge 2$): $0$,
$1,\ldots$, $N-1$. The model shall have $\mathbb{Z}_N$-symmetry and
we may extend the range of the spins $\si_{x,y}$ to all integers identifying
two values if their difference is a multiple of $N$. The model has a chiral restriction on the
values of vertically neighboring spins:
 \begin{equation}
 \label{adj-rule}
  \si_{x,y}-\si_{x,y+1}=0 \mbox{\ or\ } 1 \quad {\rm mod}\ N\ .
\end{equation}
In the following we will consider the spin variables on two adjacent rows:
$(k,\,l)$ and  $(k,\,l+1)$, where $l$ is fixed and $k=1,\ldots,n+1$.
Let us denote $\si_{k,\,l}=\g_{k}$ and $\si_{k,\,l+1}=\g^\prime_{k}$.
The model depends on the parameters $t_q$ and  $\ma_k\,, \mb_k\,,
\mc_k\,, \md_k$, $\pra_k\,,\prb_k\,, \prc_k\,, \prd_k$,
$k=1,2,\ldots,n+1$. Each square plaquette of the lattice has the
Boltzmann weight (see Fig.\ref{tri})
\be\label{bbs_weight}
 W_\tau( \g_{k-1},
\g_{k};\sip_{k-1}, \sip_{k}) =  \sum_{m_{k-1}=0}^1
\omega^{m_{k-1}(\sip_{k} - \s_{k-1})} (-\omega t_q)^{ \s_{k} -
\sip_{k}- m_{k-1}}\times \ee
\[
\qquad \qquad \qquad\qquad \times F'_{k-1}( \s_{k-1}- \sip_{k-1},
m_{k-1})
 F''_{k}( \s_{k}- \sip_{k}, m_{k-1}),
 \]
where $\omega=e^{2\pi{\rm i}/N}$, and
\[
F'_{k}(0,0)=1, \qquad F'_{k}(0,1)=-\omega t_q\,\frac{\mc_k}{\mb_k},
\qquad F'_{k}(1,0) = \frac{\md_k}{\mb_k},\qquad F'_{k}(1,1)=-\omega
\frac{\ma_k}{\mb_k},
\]
and expressions for $ F''_{k}( \s_{k}- \sip_{k}, m_{k-1})$ are
obtained from $ F'_k ( \s_{k}- \sip_{k}, m_{k})$ by substitutions:
$\ma_k\,,$ $\mb_k\,,$ $\mc_k\,,$ $\md_k \rightarrow
 \pra_k\,,$ $\prb_k\,, $ $\prc_k\,,$ $ \prd_k\,$.

We consider the periodic boundary condition: $\s_{n+2}=\s_{1}$,
$\s'_{n+2}=\s'_{1}$, where $n+1$  is the number of sites on the
lattice along the horizontal axis. The transfer-matrix of the
periodic \BBS model is $N^{n+1}\times N^{n+1}$ matrix with matrix elements
\begin{equation}
{\bf t}_{n+1}({\bg},{\bg}')=\prod_{k=2}^{n+2}W_\tau(
\s_{k-1},
   \s_{k};\sip_{k-1}, \sip_{k}),
 \label{tm}
\end{equation}
labelled by the sets of spin variables ${\bg}=\{\s_1,\s_2,\ldots,\s_{n+1}\}$
 and ${\bg}'=\{\s'_1,\s'_2,\ldots,\s'_{n+1}\}$ of two neighbour rows.

Considering $m_k$, $k=1,\ldots,n+1$, in
(\ref{bbs_weight}) as auxiliary spin variables which take the two
values $0$ and $1$, we can rewrite transfer-matrix \r{tm} in a vertex
formulation associating a statistical weight not to the plaquettes but
to vertices each of them relating four spins: $m_{k-1}$, $m_k$,
$\s_{k}$, $\sip_{k}$ (see Fig.~1). Then the weight associated with
the $k$th vertex is
\[
 \ell_k(t_q;m_{k-1}, m_k;\s_{k}, \sip_{k})=
\omega^{m_{k-1}\sip_k-m_{k}\s_k}(-\omega t_q) ^{\s_{k}-
\sip_{k}-m_{k-1}}\times
\]
\be
 \label{tau_2} \,\qquad \qquad \qquad \qquad  \times F''_{k}(
\s_{k}- \sip_{k}, m_{k-1}) F'_{k }( \s_{k }- \sip_{k }, m_{k}).
\ee
and  the transfer-matrix (\ref{tm}) can be rewritten as
\be\label{tm_1}
{\bf t}_{n+1}({\bg},{\bg}')=\sum_{m_1, \ldots,\, m_{n+1}} \prod_{k=2}^{n+2}
\ell_k(t_q;m_{k-1}, m_k;\s_{k}, \sip_{k}).
\ee

For our construction of the \BBS model eigenvectors we will
use a description of  this model
as a quantum chain model. To the each site $k$ of the quantum chain we associate
the cyclic $L$-operator \cite{Kore,BS} acting in a two-dimensional auxiliary space
\be
\label{bazh_strog}L_k(\lm)=\left( \ba{ll}
1+\lm \kp_k \hv_k,\ &  \lm \hu_k^{-1} (a_k-b_k \hv_k)\\ [3mm]
\hu_k (c_k-d_k \hv_k) ,& \lm a_k c_k + \hv_k {b_k d_k}/{\kp_k }
\ea \right),\hspace*{4mm} k=1,2,\ldots,n+1.
\ee
At each site $k$ we define ultra-local Weyl elements $\hu_k$ and $\hv_k$
obeying the commutation rules and normalization
 \be\label{comrel_uv} \bu_j
\bu_k=\bu_k \bu_j\,, \quad\;\bv_j \bv_k=\bv_k
\bv_j\,,\quad\; \bu_j
\bv_k=\om^{\delta_{j,k}}\bv_k\bu_j\,,\quad\; \om=e^{2\pi i/N},
\quad \bu_k^N=\bv_k^N=1.
\ee
In \r{bazh_strog},
$\lm$ is the spectral parameter and we have five
parameters $\:\kp_k,\;a_k,\;b_k,\;c_k,\;d_k\,$ per site. At each site $k$ we
define a $N$-dimensional linear space (quantum space) ${\cal V}_k$
with the basis $ |\g\ra_k$, $\g\in \ZN$ and natural
scalar product $\: _k\langle\g'|\g\ra_k=\delta_{\g',\g}.$ In
${\cal V}_k$ the Weyl elements $\bu_k$ and $\bv_k$ act as:
\be\label{uv} \bu_k |\g\ra_k=\om^\g |\g\ra_k\, ,\qquad
\bv_k |\g\ra_k = |\g+1\ra_k\, . \ee

The correspondence between the lattice \BBS model and its quantum chain analog is established
through the relation
\begin{equation}\label{identific}
\ell_k(t_q; m_{k-1}, m_k;\s_{k}, \sip_{k})= \, _k\langle\g_k| L_k(\lm)_{m_{k-1},m_k}|\g_k'\ra_k
\end{equation}
and the following connection between the parameters of these models
\be\label{par_rel}
\lambda=-\omega t_q\,,\quad
\kp_k=\frac{\md_k}{\mb_k} \frac{\prd_k}{\prb_k} \,,\quad
a_k=\frac{\prc_k}{\prb_k}\,,\quad b_k=\omega
      \frac{\md_k}{\mb_k}\frac{\pra_k}{\prb_k}\, ,\quad
c_k=\frac{\mc_k}{\mb_k}\,,\quad
d_k=\frac{\ma_k}{\mb_k}\frac{\prd_k}{\prb_k}\,.
\ee
We extend the action of the
operators $\bu_k$, $\bv_k$ to ${\cal V}^{(n+1)}={\cal V}_1\otimes{\cal
V}_2\otimes\cdots\otimes{\cal V}_{n+1}$ defining this action to be
trivial in all ${\cal V}_s$ with $s\ne k$.
The monodromy matrix for the quantum chain with $n+1$ sites
is defined as \be \label{mm}
{T}_{n+1}(\lm)\,=\,L_1(\lm)\,L_2(\lm)\,\cdots\, L_{n+1}(\lm)= \lk \ba{ll}
A_{n+1}(\lm)& B_{n+1}(\lm)\\
C_{n+1}(\lm)& D_{n+1}(\lm) \ea \rk. \ee  The transfer-matrix (\ref{tm_1}) is
obtained taking the trace in the auxiliary space
 \be \label{tr_matl} {\bf t}_{n+1}(\lambda)\:=\:\tr\:
T_{n+1}(\lm)\:=\:A_{n+1}(\lm)+D_{n+1}(\lm)\,.
\ee
This  quantum chain is
integrable because the $L$-operators \r{bazh_strog} are
intertwined by the twisted 6-vertex $R$-matrix
\be
{R}(\la,\nu)\;=\;\lk\begin{array}{cccc} \la-\om\nu & 0 & 0 & 0 \\
[0.3mm] 0 & \om(\la-\nu) & \la (1-\om) & 0 \\ [0.3mm] 0 & \nu (1-\om)
& \la-\nu & 0 \\ [0.3mm] 0 & 0 & 0 & \la-\om\nu \end{array}\rk\!,
\ee
\be R(\la,\nu)\,  L^{(1)}_k\, (\la) L^{(2)}_k(\nu)=
L^{(2)}_k(\nu)\, L^{(1)}_k(\la)\, R(\la,\nu), \label{rll} \ee
where
$L^{(1)}_k(\la)= L_k(\la)\otimes\mathbb{I}$,
$L^{(2)}_k(\nu)=\mathbb{I} \otimes L_k(\nu)$. Relation
(\ref{rll}) leads to  $[{\bf t}_{n+1}(\lambda),{\bf t}_{n+1}(\mu)]=0$.
So ${\bf t}_{n+1}(\la)$ is a generating
 function for the commuting set of non-local and
non-hermitian Hamiltonians of the model.
It also follows from the intertwining relation (\ref{rll}) that $[B_{n+1}(\lambda),B_{n+1}(\mu)]=0$
and therefore $B_{n+1}(\lm)$ is a generating function for another commuting set of operators.

The BBS model with fixed-spin boundary conditions can be obtained  using
periodic inhomogeneous BBS model with $n+1$ sites (in each row if we consider the lattice formulation) if one fixes
$a'_{n+1}=d'_{n+1}=0$. In this case $F'_{n+1}(\g_{n+1}-\g'_{n+1},m)=0$ unless $\g_{n+1}=\g'_{n+1}$.
It means that ${\bf t}_{n+1}(\lm)$ with such parameters and fixed $\g_{n+1}$
can be interpreted as transfer-matrix of $n$-site
BBS model with first and last ($n$th) spins interacting with fixed spin $\g_{n+1}$
(fixed-spin boundary conditions).

Let us look what we have in $L$-operator formulation. Due to \r{par_rel},
the relations $a'_{n+1}=d'_{n+1}=0$ give us $\kp_{n+1}=b_{n+1}=d_{n+1}=0$ and $b_{n+1}d_{n+1}/\kp_{n+1}=0$.
So we get
\be\label{Lnp1t}
L_{n+1}(\lm)=\left( \ba{ll}
1,\ &  \lm \hu_{n+1}^{-1} a_{n+1}\\
\hu_{n+1} c_{n+1} ,& \lm a_{n+1} c_{n+1}  \ea \right)=
\left( \ba{l}
1,\\ \hu_{n+1} c_{n+1} \ea \right)
\left( \ba{ll}
1,\ &  \lm \hu_{n+1}^{-1} a_{n+1} \ea \right)\,.
\ee
Since ${\bf t}_{n+1}(\lm)$ commutes with $\hu_{n+1}$ we may restrict the space of states of
$(n+1)$-site problem to the states with fixed eigenvalue of $\hu_{n+1}=\om^{\g_{n+1}}$ in order
to obtain $n$-site problem. To consider different fixed spins on both boundaries
 we include one more $L$-operator $L_0(\lm)$
like \r{Lnp1t} but with operator $\hu_{0}$ and
parameters $a_0$ and $c_0$. On the states with fixed eigenvalues $\hu_{n+1}=\om^{\g_{n+1}}$
and $\hu_{0}=\om^{\g_{0}}$ we have
\[
L_{n+1}(\lm)L_{0}(\lm)=(1+\lm \om^{\g_0-\g_{n+1}}a_{n+1}c_0)\cdot
\left( \ba{ll} 1,\ &  \lm   a_{0} \om^{-\g_{0}}\\
c_{n+1}\om^{\g_{n+1}} ,& \lm a_{0} c_{n+1}  \om^{\g_{n+1}-\g_{0}} \ea \right).
\]
It means that we can imitate different boundary spins by one $L$-operator
\be\label{Lnp1}
L_{n+1}(\lm)=
\left( \ba{ll} 1,\ &  \lm \hu_{n+1}^{-1} a_{0} \om^{\g_{n+1}-\g_{0}}\\
\hu_{n+1} c_{n+1} ,& \lm a_{0} c_{n+1}  \om^{\g_{n+1}-\g_{0}} \ea \right),
\ee
which is of the form \r{bazh_strog} with special parameters.
Thus in this paper we consider the monodromy matrix defined by \r{mm}
for the quantum chain with $n+1$ sites, where $L_{n+1}(\lm)$ is given by \r{Lnp1}.
The corresponding transfer-matrix with fixed boundary spins is
\be\label{tmbfs}
{\bf t}^{\rm B}(\lm)=\tr {T}_{n+1}(\lm)=\left( \ba{ll}
1,\ &  \lm \om^{-\g_{0}} a_{0} \ea \right)
L_1(\lm)\,L_2(\lm)\,\cdots\, L_n(\lm) \left( \ba{l}
1,\\ \om^{\g_{n+1}} c_{n+1} \ea \right).
\ee

The problem of construction of the eigenvectors of the transfer-matrix ${\bf t}^{\rm B}(\lm)$
will be solved as follows.
First, we find the eigenvectors of $B_1(\lm)$. Then the eigenvectors of $B_m(\lm)$ are obtained
recursively as linear combination of the eigenvectors of $B_{m-1}(\lm)$.
After that the eigenvectors for the transfer-matrix \r{tmbfs} are
constructed  as linear combinations of the eigenvectors of $B_{n+1}(\lm)$
(the auxiliary model). The multi-variable coefficients of this decomposition
admit the separation of variables and can be written as products of
single-variable functions, each satisfying a Baxter difference
equation which can be solved explicitly.

\section{Eigenvalues and eigenvectors of $B_m(\lm)$}

\subsection{Eigenvalues of $B_m(\lm)$}
\label{recinh}

We start from the problem of construction of the eigenvectors of $B_m(\lm)$.
According to $[B_m(\lambda),B_m(\mu)]=0$, we are looking for the eigenvectors of $B_m(\lm)$ not depending on
$\lm$ with the eigenvalue being a polynomial in $\lm$.
Factorizing this polynomial we get
\be\label{iBlm}
B_m(\lm)\Psi_{\bdr_m} =\lm\, r_{m,0}\,\om^{-\rho_{m,0}}
\prod_{s=1}^{m-1}\lk\lm+r_{m,s}\om^{-\rho_{m,s}}\rk\Psi_{\bdr_m}\,,
\ee
where $r_{m,s}$, $s=0,\ldots,m-1$, is a set of constants
and we shall use the phases
\be    \bdr_m=(\rho_{m,0},\ldots,\rho_{m,m-1})\in(\ZN)^m\label{br}\ee
as labels of the eigenvectors. These formulas are valid for all $m=1,\ldots,n+1$,
that is including the chain with boundary $L$-operator \r{Lnp1}. As we shall see later,
in the latter case not all the phases $\bdr_{n+1}$ are possible but only those which satisfy the
following restriction related to the boundary spin $\g_0$:
\be\label{trhoB}
\tilde \rho_{n+1}:=\sum_{s=0}^n \rho_{n+1,s}=\g_0+1\quad\mbox{mod}\,N.
\ee

Let us define the ``averaged'' counterpart
$\,{\cal O}(\lm^N)\,$ of a quantum cyclic operator $\,O(\lm)\,$ using
averaging procedure \cite{Tarasov}
\be {\cal O}(\lm^N)\;=\;\langle\,O\,\rangle(\lm^N)=\;{\textstyle \av} O(\om^s\lm)
\label{ave}\ee
and apply this procedure to the entries of the quantum $L$-operator \r{bazh_strog}.
Denote the result by $\mathcal{L}_k(\lm^N)$
\be\label{Lclass}
 \mathcal{L}_k(\lm^N)
=\left( \ba{cc}1-\epsilon \kp_k^N \lm^N &\;\; -\epsilon\lm^N(a_k^N-b_k^N)\\
            c_k^N-d_k^N &\;\; b_k^N d_k^N/\kp_k^N-\epsilon \lm^N a_k^N c_k^N\ea \right),\ee
where $\epsilon=(-1)^N$, and call it as the ``averaged'' $\mathcal{L}$-operator
of the  BBS model.
In particular, the averaging of \r{Lnp1} gives
\be\label{LclassLnp1}
 \mathcal{L}_{n+1}(\lm^N)
=\left( \ba{cc}1&\;\; -\epsilon\lm^N a_0^N\\
            c_{n+1}^N &\;\; -\epsilon \lm^N a_0^N c_{n+1}^N\ea \right).\ee
Accordingly, the averaged
monodromy ${\cal T}_m$ for the $m$-site chain is
\be \label{Tclass}
\mathcal{T}_m(\lm^N)\:=\;{\cal L}_1(\lm^N)\: {\cal L}_2(\lm^N) \cdots\:
{\cal L}_m(\lm^N)\:=\:\left(\ba{cc}
{\cal A}_m(\lm^N) &  {\cal B}_m(\lm^N)\\
{\cal C}_m(\lm^N) &  {\cal D}_m(\lm^N)
\ea\right),\ee
where the entries are polynomials of $\lm^N$.
 By Proposition~1.5 from \cite{Tarasov} (see also \cite{Roan}), these
polynomials coincide with averages $\langle{A}_m\rangle$, $\langle{B}_m\rangle$,
$\langle{C}_m\rangle$ and $\langle{D}_m\rangle$ of the entries of \r{mm}.
This proposition provides a tool for finding the $N$-th powers of the amplitudes $r_{m,s}$:
applying \r{ave} to \r{iBlm} we obtain
\be
 {{\cal B}_m(\lm^N)=(-\epsilon)^m \lm^N r_{m,0}^N \prod_{s=1}^{m-1}(\lm^N-
\epsilon\, r_{m,s}^N)\,.}\label{Brec}
\ee
This relation together with \r{Lclass} and \r{Tclass} allows to find $r_{m,s}^N$ in terms of
the parameters $a_k^N$, $b_k^N$, $c_k^N$, $d_k^N$ and $\kp_k^N$, $k=1,\ldots,m$.
The problem of finding the amplitudes $r_{m,s}$ is reduced
to the problem of solving a $(m-1)$-th degree algebraic relation.
As shown in \cite{GIPS}, in
the case of the homogeneous \BBS chain model
the problem is reduced to solving a quadratic equation only.
The described procedure gives the amplitudes $r_{m,s}$ up to some roots of unity.
In fact we can fix these phases arbitrarily because this leads just to relabeling of the
eigenvectors. In what follows we suppose that we fixed some solution
$\{r_{m,s}\}$ in terms of the parameters $a_k^N$, $b_k^N$, $c_k^N$, $d_k^N$ and $\kp_k^N$.
Again in the case $m=n+1$ we can not define the phase of $r_{n+1,0}$ arbitrarily. We fix it by the relation
\be\label{trB}
\tilde r_{n+1}:=\prod_{s=0}^n r_{n+1,s}=\om a_0.
\ee
Then the relations \r{trhoB} and \r{trB} provide correct coefficient $a_0\om^{-\g_0}$
at $\lm$ (the lowest term) in the eigenvalue of $B_{n+1}(\lm)$.

\subsection{One-site eigenvectors for the auxiliary problem}

In BBS model a very important role is played by the
cyclic function $w_p(\g)$ \cite{BB} which
depends on a $\ZN$-variable $\g$ and on a point $p=(x,y)$
restricted to the Fermat curve $x^N+y^N=1$. We define $w_p(\g)$ by the difference equation
\[
\frac{w_p(\g)}{w_p(\g-1)}=\frac{y}{1-\om^\g\,x}\,;\qquad
w_p(0)=1\,; \qquad \g\in\ZN.
\]
The Fermat curve restriction guarantees the cyclic property $w_{p}(\g+N)=w_{p}(\g).$
The function $w_p(\g)$ is a root of unity analog of the $q$-gamma function.

It is convenient to change the bases in the spaces ${\cal V}_k$.
Instead of $|\g\ra_k$, $\g\in \ZN$, we will use the vectors
\be\label{psik}
\psi^{(k)}_{\rho_k}=
\sum_{\g\in \ZN} w_{p_k}(\g-\rho_k)|\g\ra_k\,,\qquad \rho_k\in \ZN\,,
\ee
which are eigenvectors of the upper off-diagonal matrix element
of the operator $L_k(\lm)$:
\be  \la\,\bu_k^{-1} (a_k-b_k \bv_k)\, \psi^{(k)}_{\rho_k}
=\la\,r_k\,\om^{-\rho_k}\,\psi_{\rho_k}^{(k)}.
  \label{efb} \ee
The coordinates of the Fermat curve points
$p_k=(x_k,y_k)$ are defined as follows.
Let us fix some value of $r_k$ to satisfy $r_k^N=a_k^N-b_k^N$.
Then
\be x_k=r_k/a_k,\qquad y_k=b_k/a_k\,.\label{xyab} \ee

In the case $k=n+1$ that is for \r{Lnp1}, the eigenvectors \r{psik} degenerate as follows.
We fix $x_{n+1}=\om^{\g_0-\g_{n+1}}$, $y_{n+1}=0$, $r_{n+1}=a_0$.
It gives, up to an inessential constant multiplier, $w_{p_{n+1}}(\g)=\delta_{\g,\g_{n+1}-\g_0}$ and therefore
$\psi^{(n+1)}_{\rho_{n+1}}=$ $|\rho_{n+1}+\g_{n+1}-\g_0\ra_{n+1}$. The vector $\psi^{(n+1)}_{\rho_{n+1}}$
satisfies \r{efb} with $k=n+1$. On the states with fixed eigenvalue
$\hu_{n+1}=\om^{\g_{n+1}}$ we have to identify $\rho_{n+1}=\g_0$.

 Note, if $r_k=0$ (in particular, in the superintegrable
 case) it leads to $x_k=0$, $y_k=1$. In this case the cyclic  function $w_p(\gamma)$
degenerates and \r{psik} does not give a new basis in ${\cal V}_k$.
So we consider the generic parameters such that $r_k\ne 0$.

The operator $\bv_k$ acts as cyclic ladder operator:
\be \bv_k\psi_{\rho_k}^{(k)}= \psi_{\rho_k+1}^{(k)}\,.
\label{vpsi}\ee

Using \r{efb} for $k=1$ and comparing to (\ref{iBlm}), we write one-site
eigenvector as $\Psi_{\rho_{1,0}}:= \psi^{(1)}_{\rho_{1,0}}$. With $\;r_{1,0}\,=r_1\:$
we have  \be\label{AB1}
B_1(\lm)\Psi_{\rho_{1,0}}=\lm\: r_{1,0}\: \om^{-\rho_{1,0}}\, \Psi_{\rho_{1,0}}\,,
\quad
A_1(\lm)\Psi_{\rho_{1,0}}=\Psi_{\rho_{1,0}}+\lm\kp_1 \Psi_{\rho_{1,0}+1}\,.
\ee

\subsection{Fermat curve points in the formulas for
                      the eigenvectors of $B_m(\lm)$}

The formula for the eigenvectors of $B_m(\lm)$ is defined in terms
of $w_p(\g)$ function depending on four types of points on the Fermat curve $x^N+y^N=1$:
\be   \tilde p_m=(\tilde
x_m,\tilde y_m);\quad
p_{m,s}=(x_{m,s}, y_{m,s});\quad\tilde p_{m,s}=(\tilde
x_{m,s},\tilde y_{m,s});\quad
p^{m,s}_{m',s'}=(x^{m,s}_{m',s'},y^{m,s}_{m',s'}).\label{pxy}\ee
The coordinates of these points are expressed in the terms of amplitudes
$r_{m,s}$, $m=1,\ldots,$ $n+1$, $s=0,\ldots,$ $m-1$ (defined as {\it some} solutions of
equations \r{Brec}, $m=1,\ldots,$ $n+1$) by
\be  x^{m,s}_{m',s'}=r_{m,s}/r_{m',s'},\quad x_{m,s}=a_m \kp_m
r_{m,s}/b_m,\quad \tilde x_{m,s}=d_m/(\kp_m c_m r_{m,s}),\quad s,s'\ge 1.
\label{xxx}\ee
In the construction of the eigenvectors of $B_m(\lm)$ we need the points of
type $p_{m,s}$ and $\tilde p_{m,s}$, $s\ge 1$, only for $m\le n$.
For all the other types of points $m$ runs up to $n+1$.

The values of $y^{m,s}_{m',s'}$, $y_{m,s}$, $\tilde y_{m,s}$ are defined
by the condition on $p^{m,s}_{m',s'}$, $p_{m,s}$, $\tilde p_{m,s}$ to belong to the Fermat curve
and the following relations ($1\le l\le m-2$) on phases of $y$-coordinates:
\be \label{rel_other}
\frac{\tilde r_{m-1} r_{m,0}\,r_{m-1}}{\tilde r_{m-2}\,r_{m-1,0} r_{m}\,
b_{m-1}\,c_{m-1}\, y_{m-1,l}\,\tilde y_{m-1,l}}\;
 \prod_{s\ne l}^{m-2} \frac{y^{m-1,l}_{m-1,s}}{y^{m-1,s}_{m-1,l}}
\;\frac{\prod_{k=1}^{m-1} \;y^{m,k}_{m-1,l}}{\prod_{s=1}^{m-3}\; y_{m-2,s}^{m-1,l}}=1\,,
\ee
where
\be \rt_m\,=\,r_{m,0}\,r_{m,1}\,\ldots\, r_{m,m-1}\,.\label{rtilde}\ee
Practically we may define $y^{m,s}_{m',s'}$, $y_{m,s}$ arbitrarily and the coordinates
$\tilde y_{m,s}$ by \r{rel_other}. Then the points $\tilde p_{m,s}$, due to \r{Brec}, will belong
to the Fermat curve automatically \cite{GIPS}.

The coordinates of the points $p_{m,0}$ and ${\tilde p}_m$, $1\le m\le n$, are defined by
\be x_{m,0} r_{m,0}=r_{m-1,0} a_m c_m,\qquad y_{m,0}
r_{m,0}=\kp_1\kp_2\cdots\kp_{m-1} r_m\ ,\label{xynull}\ee
\be \tilde x_m \tilde r_m =r_m,\qquad \tilde y_m \tilde r_m =b_m
d_m \tilde r_{m-1}/\kp_m\,.\label{xyt}\ee

We use the same formulas for the case $m=n+1$. Due to \r{trB} and $r_{n+1}=a_0$,
the solution of \r{xyt} is $\tilde x_{n+1}=\om^{-1}$, $\tilde y_{n+1}=0$. Correct limiting procedure gives
$(w_{\tilde p_m}(\rho))^{-1}=\delta_{\rho,0}$. We define
the coordinates of the Fermat curve point $p_{n+1,0}$ by
\be x_{n+1,0} r_{n+1,0}=r_{n,0} a_0 c_{n+1},\qquad y_{n+1,0}
r_{n+1,0}=\kp_1\kp_2\cdots\kp_{n} a_0\,. \label{xynullB}\ee
Note that from \r{xynull} we had to define $x_{n+1,0}$ by
relation $x_{n+1,0} r_{n+1,0}=r_{n,0} a_0 c_{n+1} \om^{\g_{n+1}-\g_0}$ but we want to avoid entering the
values of boundary spins into the definition of Fermat curve points.
Such change of definition of $x_{n+1,0}$ means that instead of
$w_{(x_{n+1,0},y_{n+1,0})}(\rho)$ we have to use
\be\label{sr}
w_{(x_{n+1,0}\om^{\g_{n+1}-\g_0},y_{n+1,0})}(\rho)=
\frac{w_{p_{n+1,0}}(\rho+\g_{n+1}-\g_0)}{w_{p_{n+1,0}}(\g_{n+1}-\g_0)}
\ee
with point $p_{n+1,0}$ defined by \r{xynullB}. This formula allows to move the dependence
on the values of boundary spins $\g_0$ and $\g_{n+1}$ to the argument of $w_{(x_{n+1,0},y_{n+1,0})}(\rho)$.

\subsection{Recursive formulas for the eigenvectors of $B_m(\lm)$}
\label{eigenvectors}

Recall from (\ref{AB1}) that the vector
$\Psi_{\rho_{1,0}}:= \psi^{(1)}_{\rho_{1,0}}\in {\cal V}_1$ is
eigenvector for $B_1(\lm)$ and
from \r{iBlm}, \r{br} that the eigenvectors $\Psi_{\bdr_m}$ of $B_m(\la)$
were labeled by the vector $\bdr_m=(\rho_{m,0},\ldots,\rho_{m,m-1})\in(\ZN)^m$.
Let us further define:
\be  \trh_m={\textstyle \sum_{k=0}^{m-1}}\;\;\rho_{m,k};\qquad
\trh'_m={\textstyle \sum_{k=1}^{m-1}}\;\;\rho_{m,k};\qquad
\bdr_m'=
(\rho_{m,1},\ldots,\rho_{m,m-1})\in(\ZN)^{m-1}. \ee
The vector $\bdr_m^{\pm k}\;$ denotes the vector $\bdr_m$ in which $\rho_{m,k}$ is
replaced by $\rho_{m,k}\pm 1$, i.e.
\[
\hs\hs\hx\;\:\bdr_m^{\pm k}=(\rho_{m,0},\ldots,\rho_{m,k}\pm 1,\ldots,\rho_{m,m-1}),\hq
k=0,1,\ldots,m-1.
\]

The following Theorem~\ref{th1} proved in \cite{GIPS} gives a recursive procedure to obtain the eigenvectors
$\Psi_{\bdr_m}\:\in\:{\cal V}^{(m)}$, $2\le m\le n$,
of $B_m(\lm)$ from the eigenvectors
$\Psi_{\bdr_{m-1}}\in\,{\cal V}^{(m-1)}$ of $B_{m-1}(\lm)$
and single site vectors $\psi^{(m)}_{\rho_m}\:\in\:{\cal V}_{m}$
defined by (\ref{psik}). The recursion is starting from the already defined $\Psi_{\rho_{1,0}}$.
The Theorem~1 is valid provided  $r_m^N\ne 0$,
the polynomials ${\cal B}_m(\lm^N)/\lm^N$, $m=2,\ldots,n$,
have nonzero simple zeros and $\det {\cal T}_n(\epsilon r_{m,s}^N)\ne 0$
(cf. the definition of the $B$-representation in \cite{Tarasov}).

\begin{theorem}\label{th1}
The vector
\be
\Psi_{\bdr_m}=\sum_{\bdr_{m-1}\in (\ZN)^{m-1}\atop \rho_m \in \ZN}
 Q(\bdr_{m-1},\rho_m|\bdr_m) \Psi_{\bdr_{m-1}}\otimes
\psi^{(m)}_{\rho_m} \label{PSI}\ee
where
\bea
Q(\bdr_{m-1},\rho_m|\bdr_{m})&=&\frac{\om^{(\tilde \rho_m-
\tilde \rho_{m-1}) (\rho_m-\rho_{m,0})}}
{w_{p_{m0}}(\rho_{m,0}-\rho_{m-1,0}-1) w_{\tilde p_m}(\tilde
\rho_m- \rho_m-1)}\: \times \ny  \\\label{QQQ} &&\hspace*{-2cm}\times\:
\frac{ \prod_{l=1}^{m-2} \prod_{k=1}^{m-1}
w_{p_{m-1,l}^{m,k}}(\rho_{m-1,l}-\rho_{m,k})} {\prod_{j,l=1\atop
(j\ne l)}^{m-2} w_{p_{m-1,j}^{m-1,l}}(\rho_{m-1,j}-\rho_{m-1,l})}
\prod_{l=1}^{m-2} \frac{w_{p_{m-1,l}}(-\rho_{m-1,l})}{w_{\tilde
p_{m-1,l}}(\rho_{m-1,l})}\eea
is eigenvector of $B_m(\la)$:
\be\label{Blm}
B_m(\lm)\Psi_{\bdr_m} =\lm\, r_{m,0}\om^{-\rho_{m,0}}
\prod_{k=1}^{m-1}\lk\lm+r_{m,k}\om^{-\rho_{m,k}}\rk\Psi_{\bdr_m}.
\ee

At the $m-1$ zeros $\la_{m,k}$ of the
eigenvalue polynomial of $B_m(\la)$
\be \la_{m,k}=-r_{m,k}\om^{-\rho_{m,k}},\hs\hs k=1,\ldots,m-1,\ee
the operators $A_m$ and $D_m$ act as shift operators for the $k$-th (and $0$-th for $D_m$)
index of $\Psi_{\bdr_m}$:
\be
 A_m\lk\lm_{m,k}\rk\Psi_{\bdr_m}= \varphi_k(\bdr_m')\;\Psi_{\bdr_m^{+k}}\,,\label{Almk}
\qquad D_m(\lm_{m,k})\Psi_{\bdr_m}=\tilde\varphi_k(\bdr_m')\,\Psi_{\bdr_m^{+0,-k}}\,, \label{Dlmk}
\ee
where
\be \label{phik} \varphi_k(\bdr_m')\;=\;-\frac{\tilde r_{m-1}}{r_m}\;\om^{-\tilde
\rho'_{m}}\;F_m(\lm_{m,k}/\om)\;\prod_{s=1}^{m-2}y_{m-1,s}^{m,k}\,,
 \ee
 \be
  \tilde\varphi_k(\bdr'_m)=
-\frac{r_m}{\tilde r_{m-1}}\frac{\om^{\trh'_m-1}}
{\prod_{s=1}^{m-2}y_{m-1,s}^{m,k}}\;\prod_{s=1}^{m-1}\;F_s(\lm_{m,k})\,,
 \ee
 \be F_s(\lm)\:=\:\lk\, b_s\,+\om a_s \,\kp_s \lm\rk\,
\lk \,\la\, c_s\,+d_s/\kp_s\, \rk. \label{qdet}\ee
\end{theorem}

The next theorem gives the formula for the eigenvectors of $B_{n+1}(\lm)$ in the space ${\cal V}^{(n)}$
(which is restriction of the initial space of states ${\cal V}^{(n+1)}$ to the subspace of
fixed value $\g_{n+1}$ of spin at $(n+1)$th site) in terms of the eigenvectors of $B_{n}(\lm)$.
As was explained before, the components of $\bdr_{n+1}$ are not independent but satisfy
\r{trhoB}. So we will label the eigenvectors $\Psi^{\rm B}_{\bdr'_{n+1}}$ of $B_{n+1}(\lm)$
by the set $\bdr_{n+1}'=(\rho_{n+1,1},\ldots,\rho_{n+1,n})$. Then
$\rho_{n+1,0}=\g_0+1-\tilde\rho'_{n+1}$.
\begin{theorem}\label{th2}
The vector
\be
\Psi^{\rm B}_{\bdr'_{n+1}}=\sum_{\bdr_{n}\in (\ZN)^{n}}
 Q^{\rm B}(\bdr_{n}|\bdr'_{n+1}) \Psi_{\bdr_{n}} \label{PSIB}\ee
where
\bea
Q^{\rm B}(\bdr_{n}|\bdr'_{n+1}) &=&\frac{\om^{(\g_0+1- \tilde \rho_{n}) (\tilde\rho'_{n+1}-1)}}
{w_{p_{n+1,0}}(\g_{n+1}-\tilde\rho'_{n+1}-\rho_{n,0})} \: \times \ny  \\\label{QQQB} &&\hspace*{-2cm}\times\:
\frac{ \prod_{l=1}^{n-1} \prod_{k=1}^{n}
w_{p_{n,l}^{n+1,k}}(\rho_{n,l}-\rho_{n+1,k})} {\prod_{j,l=1\atop
(j\ne l)}^{n-1} w_{p_{n,j}^{n,l}}(\rho_{n,j}-\rho_{n,l})}
\prod_{l=1}^{n-1} \frac{w_{p_{n,l}}(-\rho_{n,l})}{w_{\tilde
p_{n,l}}(\rho_{n,l})}\eea
is eigenvector of $B_{n+1}(\la)$:
\be\label{BlmB}
B_{n+1}(\lm)\Psi^{\rm B}_{\bdr'_{n+1}} =\lm\, r_{n+1,0}\om^{-\rho_{n+1,0}}
\prod_{k=1}^{n}\lk\lm+r_{n+1,k}\om^{-\rho_{n+1,k}}\rk\Psi^{\rm B}_{\bdr'_{n+1}}\,.
\ee

At the $n$ zeros $\la_{n+1,k}$ of the
eigenvalue polynomial of $B_{n+1}(\la)$
\be \la_{n+1,k}=-r_{n+1,k}\om^{-\rho_{n+1,k}},\hs\hs k=1,\ldots,n,\ee
the operators $A_{n+1}$ and $D_{n+1}$ act as:
\be
 A_{n+1}\lk\lm_{n+1,k}\rk\Psi^{\rm B}_{\bdr'_{n+1}}=0\,,\label{AlmkB}
\qquad D_{n+1}(\lm_{n+1,k})\Psi^{\rm B}_{\bdr'_{n+1}}=
\tilde\varphi_k(\bdr_{n+1}')\,\Psi^{\rm B}_{\bdr_{n+1}^{'-k}}\,. \label{DlmkB}
\ee
\end{theorem}

{\it Proof}.
The formula for the eigenvectors $\Psi^{\rm B}_{\bdr'_{n+1}}$ of $B_{n+1}(\lm)$ follows from the formulas
\r{PSI} and \r{QQQ} at $m=n+1$ after appropriate limiting procedure:
the factor $w_{\tilde p_{n+1}}(\tilde \rho_{n+1}- \rho_{n+1}-1)$ kills summation over $\rho_{n+1}$
(fixing $\rho_{n+1}=\g_0$) and becomes $1$,
the factor $w_{p_{n+1,0}}(\rho_{n+1,0}-\rho_{n,0}-1)$ due to \r{sr} becomes
$w_{p_{n+1,0}}(\g_{n+1}-\tilde\rho'_{n+1}-\rho_{n,0})$
up to an inessential constant multiplier $w_{p_{n+1,0}}(\g_{n+1}-\g_0)$.
The action formula for $A_{n+1}$ follows from the identity $\lm a_0 \om^{-\g_0} A_{n+1}(\lm)=B_{n+1}(\lm)$.
\hfill $\Box$

\section{Eigenvalues and eigenvectors of the boundary\\ transfer-matrix ${\bf t}^{\rm B}(\lm)$}
\label{periodic}

After having determined in Theorem~2 the eigenvectors $\Psi^{\rm B}_{\bdr'_{n+1}}$
of the auxiliary system, we are looking for the eigenvectors of ${\bf t}^{\rm B}(\lm)$ from \r{tmbfs} as
linear combinations of the eigenvectors $\Psi^{\rm B}_{\bdr'_{n+1}}$.

From the expressions for the $L$-operators we can see
that the eigenvalue of ${\bf t}^{\rm B}(\lm)$ is
\be  t^{\rm B}(\lm|\g_0,\,\g_{n+1},\,{\bf E})\;=\;
E_0+E_1\lm+\cdots+E_{n}\lm^{n}+E_{n+1}\lm^{n+1}\,, \label{tnh}\ee
where ${\bf E}=\{E_1,\ldots,E_{n}\}$ and the values of $E_0$ and $E_{n+1}$
are
\be\label{iE0En}
 E_0\,=1,\qquad
E_{n+1}\,=\prod_{m=1}^n a_m c_m\cdot a_0 c_{n+1}\om^{\g_{n+1}-\g_0}\,.\ee
The possible values of ${\bf E}$ can be found from the functional relations.
We define the following polynomials depending on unknown set ${\bf E}$:
 $\tau^{(0)}(\lm)=0$,
$\tau^{(1)}(\lm)=1$, $\tau^{(2)}(\lm)=t^{\rm B}(\lm)$  and recursively
\be \tau^{(j+1)}(\lm)=\tau^{(2)}(\om^{j-1}\lm)\,\tau^{(j)}(\lm)-
\om^{\rho}\,z(\om^{j-1}\lm)\, \tau^{(j-1)}(\lm),\qquad j=2,3,\ldots,
N, \label{rectau} \ee
where $z(\lm)=\prod_{m=1}^{n+1} F_m(\lm/\om)$.
Then the ``truncation'' identity
\be\label{reltau}
\tau^{(N+1)}(\lm)-
\om^{\rho}\,z(\lm)\, \tau^{(N-1)}(\om\lm)=
{\cal A}_{n+1}(\lm^N)+{\cal D}_{n+1}(\lm^N)
\ee
defines possible values of ${\bf E}$.
Note, the polynomial ${\cal A}_{n+1}(\lm^N)+{\cal D}_{n+1}(\lm^N)$ corresponds
to $\alpha_q+\bar \alpha_q$ in \cite{B_tau}.

As was mentioned in \cite{B_tau}, in the case of parameters considered in the present paper
we have $z(\lm)=0$ (due to $F_{n+1}(\lm)=0$) and therefore all the set of functional relations reduces to one relation
(we use the averaging \r{ave}):
\be\label{fr}
\langle t^{\rm B}\rangle (\lm^N)={\cal A}_{n+1}(\lm^N)+{\cal D}_{n+1}(\lm^N)\,
\ee
or, equivalently, $\langle A_{n+1}+D_{n+1}\rangle =\langle A_{n+1}\rangle+\langle D_{n+1}\rangle$.
At this point it is natural to conjecture the formula for the eigenvalues of transfer-matrix \cite{B_tau}:
\be\label{evaltm}
 t^{\rm B}(\lm|{\bm \sigma}_{n+1})=\prod_{j=1}^{n+1}\left(\frac{\lm}{s_j\om^{-\sigma_j}}+1\right),
 \ee
where ${\bm \sigma}_{n+1}=(\sigma_1,\sigma_2,\ldots,\sigma_{n+1})\in(\mathbb{Z}_N)^{n+1}$ and
\be\label{rel_sigma}
\sigma_1+\sigma_2+\cdots+\sigma_{n+1}=\g_{n+1}-\g_0\quad\mbox{mod}\,N,
\ee
The numbers $s_j$, $j=1,2,\ldots,n+1$, satisfy two relations:
\be\label{relprods}
\prod_{j=1}^{n+1}{s_j}^{-1}= \prod_{m=1}^n a_m c_m\cdot a_0 c_{n+1},
\ee
\be\label{rel_s}
\prod_{j=1}^{n+1}\left(1-(-1)^N \frac{\lm^N}{s_j^N}\right)={\cal A}_{n+1}(\lm^N)+{\cal D}_{n+1}(\lm^N).
\ee
The relation \r{rel_s} follows from functional relation \r{fr} and averaged \r{evaltm}.
The relation \r{rel_s} means that $(-1)^Ns_j^N$ are roots of polynomial ${\cal A}_{n+1}(\lm^N)+{\cal D}_{n+1}(\lm^N)$
with respect to $\lm^N$.
Note, the relations \r{rel_sigma} and  \r{relprods} provide correct value of $E_{n+1}$ in
\r{iE0En}.

In the paper \cite{B_tau}, Baxter conjectured  that to each set
${\bm \sigma}_{n+1}$ satisfying \r{rel_sigma}
there corresponds the eigenvector $\Phi_{{\bm \sigma}_{n+1}}$ of transfer-matrix ${\bf t}^{\rm B}(\lm)$
with eigenvalue \r{evaltm}.
The method of functional relations gives in principle the eigenvalues without information
on their multiplicities ($0$,$1$ or more). The method of separation of variables developed in \cite{GIPS} for
BBS model gives the formulas for the eigenvectors if the eigenvalues are provided. In particular, more
precise information on the multiplicities of the eigenvalues can be obtained.

According to \cite{GIPS}, we
are looking for $\Phi_{{\bm \sigma}_{n+1}}$ to be of the form
\be \Phi_{{\bm \sigma}_{n+1}}=
\sum_{\bdr'_{n+1}}  Q^{\rm B}(\bdr'_{n+1}|{\bm \sigma}_{n+1})\:\Psi^{\rm B}_{\bdr'_{n+1}},
\label{EV}\ee
where
\[ Q^{\rm B}(\bdr'_{n+1}|{\bm \sigma}_{n+1})=\frac{\prod_{k=1}^{n}\: \tilde
q_k(\rho_{n+1,k})} {\prod_{m,m'=1\atop (m\ne m')}^{n}
w_{p_{n+1,m}^{n+1,m'}}(\rho_{n+1,m}-\rho_{n+1,m'})}\]
and the functions of separated variables $\qt_k(\rho_{n+1,k})$ satisfy the Baxter type difference equations,
$k=1,2,\ldots,n$:
\be
 t^{\rm B}(\lm_{n+1,k}|{\bm \sigma}_{n+1})\;\qt_k(\rho_{n+1,k})\;=\;
\Delta_k^+(\lm_{n,k})\;\qt_k(\rho_{n+1,k}+1)\;+\;\Delta_k^-(\om\lm_{n+1,k})\;
\qt_k(\rho_{n+1,k}-1)\label{BAX}\ee
with
\be \Delta^+_k(\lm)=\chi_k^{-1}\,(\lm/\om)^{-n}\:
         \prod_{m=1}^{n}\,F_m(\lm/\om);\qquad
\Delta^-_k(\lm)=\chi_k\:(\lm/\om)^{n}\:F_{n+1}(\lm/\om); \label{Dpm}\ee
\be
\chi_k\;=\;\frac{r_{n+1,0}\:\rt_{n}}{a_0\:\rt_{n+1}}\:(\prod_{m=1\atop m\ne
k}^{n}\; y_{n+1,k}^{n+1,m}/y^{n+1,k}_{n+1,m})\:\prod_{m=1}^{n-1}\:y^{n+1,k}_{n,m}\,.
\label{chi} \ee
Since in the case of parameters considered in this paper we have $F_{n+1}(\lm)=0$ we get
$\Delta_k^-(\lm)=0$, and equations \r{BAX} can be solved explicitly in terms of functions $w_p(\g)$.
The answer for  $Q^{\rm B}(\bdr'_{n+1}|{\bm \sigma}_{n+1})$ is
\be\label{QBB}
Q^{\rm B}(\bdr'_{n+1}|{\bm \sigma}_{n+1})=\frac{\prod_{k=1}^{n} \prod_{j=1}^{n+1}
w_{p^{\rm BB}_{j,k}}(\sigma_{j}-\rho_{n+1,k})} {\prod_{m,m'=1\atop (m\ne m')}^{n}
w_{p_{n+1,m}^{n+1,m'}}(\rho_{n+1,m}-\rho_{n+1,m'})}\cdot
\prod_{k=1}^n \prod_{m=1}^n \frac{w_{p^{\rm B}_{m,k}}(\rho_{n+1,k}-1)}
{w_{\tilde p^{\rm B}_{m,k}}(-\rho_{n+1,k}-1)}\,,
\ee
where the Fermat curve points $p^{\rm BB}_{j,k}$, $p^{\rm B}_{m,k}$ and $\tilde p^{\rm B}_{m,k}$ are defined by
\[
x^{\rm BB}_{j,k}=r_{n+1,k}/s_j,\qquad x^{\rm B}_{m,k}=b_m/(a_m \kp_m r_{n+1,k}),\qquad
\tilde x^{\rm B}_{m,k}=r_{n+1,k} c_m\kp_m/d_m
\]
and conditions on the discrete phases of $y^{\rm BB}_{j,k}$, $y^{\rm B}_{m,k}$,
$\tilde y^{\rm B}_{m,k}$ :
\[
\prod_{m=1}^n \om a_m d_m y^{\rm B}_{m,k} \tilde y^{\rm B}_{m,k} = \chi_k \prod_{j=1}^n y^{\rm BB}_{j,k}\,,
\qquad k=1,2,\ldots,n\,.
\]
The validity of these relations up to $N$th root of unity is provided by \r{rel_s}. Straightforward calculation
gives
\[
{\bf t}^{\rm B}(\lm)\Phi_{{\bm \sigma}_{n+1}} = t^{\rm B}(\lm|{\bm \sigma}_{n+1}) \Phi_{{\bm \sigma}_{n+1}},
\]
where the eigenvectors $\Phi_{{\bm \sigma}_{n+1}}$ and corresponding eigenvalues
are defined by \r{EV},\r{QBB},\r{evaltm}.

\section{Ising-like $\mathbb{Z}_N$-model with fixed boundary spins}

In this section we fix special values of parameters $b_k=d_k=0$ of $L$-operators \r{bazh_strog}:
\be
\label{bbs_spec}L_k(\lm)=\left( \ba{ll}
1+\lm \kp_k \hv_k,\ &  \lm \hu_k^{-1} a_k\\ [3mm]
\hu_k c_k ,& \lm a_k c_k
\ea \right),\hspace*{4mm} k=1,2,\ldots,n,
\ee
and consider the transfer-matrix \r{tmbfs} of the BBS model with fixed boundary spins.
The coefficients of this transfer-matrix expansion in $\lm$ give
a set of commuting operators. In the previous section we gave the formula for the common eigenvectors of this set.
The coefficient at $\lm$ gives Hamiltonian of
Ising-like $\mathbb{Z}_N$ quantum chain model with fixed boundary spins \cite{BaxInv,BaxterZN}
\be\label{Baxter_Ham}
{\bf H}_1=\sum_{k=1}^n \kp_k \bv_k + \sum_{k=0}^{n+1} a_k c_{k+1} \bu_k^{-1} \bu_{k+1},
\ee
where $\bu_0=\om^{\g_0}$, $\bu_{n+1}=\om^{\g_{n+1}}$ are fixed boundary spins.
The corresponding eigenvalue
\[
E_1=\sum_{j=1}^{n+1} s_j^{-1} \om^{\sigma_j}
\]
follows from \r{evaltm}, the set ${\bm \sigma}_{n+1}$ satisfy \r{rel_sigma}, and the amplitudes
$s_j$ have to be found from \r{relprods} and \r{rel_s}. Using the explicit expressions for
averaged $L$-operators \r{Lclass} with  $b_k=d_k=0$ and \r{LclassLnp1}
we can rewrite right-hand side of \r{rel_s} as
\[
{\cal A}_{n+1}(\lm^N)+{\cal D}_{n+1}(\lm^N)=
\tr \mathcal{T}_{n+1}(\lm^N)= \tr \tilde \mathcal{L}_1(\lm^N)\cdots \tilde \mathcal{L}_{n+1}(\lm^N),
\]
where
\[\tilde  \mathcal{L}_k(\lm^N) =
\left ( \ba{cc}1&0\\
            0&c_k^{-N}\ea \right)
\mathcal{L}_k(\lm^N) \left ( \ba{cc}1&0\\
            0&c_{k+1}^{N}\ea \right)
=\left( \ba{cc}1-\epsilon \lm^N \beta_{2k}^2 & -\epsilon\lm^N \beta_{2k+1}^2\\
            1& -\epsilon\lm^N \beta_{2k+1}^2\ea \right)=
\]\[
=\left( \ba{cc}1 &-\epsilon \lm^N \beta_{2k}^2 \\
            1& 0\ea \right)
\left( \ba{cc}1 & -\epsilon\lm^N \beta_{2k+1}^2\\
            1& 0\ea \right)\,
\]
with $\epsilon=(-1)^N$, $\beta_{2k+1}=(a_k c_{k+1})^{N/2}$, $\beta_{2k}=\kp_k^{N/2}$. Finally we get
\[
\tr \mathcal{T}_{n+1}(\lm^N)=(1,-\epsilon \lm^N \beta_1^2) \left\{
\prod_{k=1}^n \left( \ba{cc}1 &-\epsilon \lm^N \beta_{2k}^2 \\
            1& 0\ea \right)
\left( \ba{cc}1 & -\epsilon\lm^N \beta_{2k+1}^2\\
            1& 0\ea \right)\right\} \left( \ba{c}1 \\  1\ea \right),
\]
where the multipliers in the product over increasing $k$ are ordered from the left to the right.
Rewriting it as determinant of $(2n+2)\times(2n+2)$ tri-diagonal matrix we get $s_1^N,\ldots$, $s_{n+1}^N$ from
the eigenvalue problem for $(2n+2)\times(2n+2)$ bi-diagonal matrix
\[B=\lk\ba{cccccc} 0\;&\beta_1&0&\ldots&0&0\\
 \beta_1&0&\beta_2&\ldots&0&0\\
0&\beta_2&0&\ldots&0&0\\
 \multicolumn{6}{c}{\ldots \qquad \ldots\qquad \ldots}\\
0&0&0&\ldots&0&\beta_{2n+1} \\
 0&0&0&\ldots&\beta_{2n+1}&0 \ea\rk\,,
\]
which has $2n+2$ eigenvalues $\pm s_j^{-N/2}$, $j=1,2,\ldots,n+1$. The method of finding eigenvalues
of Hamiltonian \r{Baxter_Ham} in terms of eigenvalues of $B$ was given in \cite{BaxterZN}.

Another model with $A_n(\lm)$ as generation function of the Hamiltonians has
similar boundary conditions and was considered in \cite{Iorgov}.
A wide class of models with integrable boundary conditions can be obtained
in the reflection equation approach with the use of the cyclic $L$-operators \r{bazh_strog}.
So it is interesting to construct the eigenvectors in these models. In particular case
 of the relativistic Toda chain at root of unity,
 the formulas for the eigenvectors were obtained in \cite{IRST}.

\section*{Acknowledgements}
The authors are thankful to the organizers of the International Workshop
``Classical and Quantum Integrable Systems'' for their kind hospitality and
excellent workshop.
This work was partially supported by the grant
INTAS-05-1000008-7865 and the Ukrainian State Foundation for Fundamental Research. The
work of V.N.S. was also partially supported  by the SCOPES-project
IB7320-110848 of Swiss NSF.

\bibliographystyle{amsplain}

\end{document}